# Comparative performance analysis of AntHocNet and DSR in optimal path route determination

## T.R. Gopalakrishnan Nair[a], Kavitha Sooda[b, *], Kavya B.S[c], Sushma M[d]

[a]ARAMCO Endowed Chair- Technology, PMU, KSA,
Advanced Networking Research Group, VP, Research, Research and Industry Incubation Center (RIIC), Dayananda Sagar Institutions,Bangalore - 560078, India
[b]Advanced Networking Research Group (RIIC), DSI,
and Asst. Professor, Dept. of CSE, Nitte Meenakshi Institute of Technology, Bangalore - 560064, India
[c,d]Lecturer, Dept. of CSE, Nitte Meenakshi Institute of Technology, Bangalore - 560064, India

**Abstract**

One of the key challenges in current Internet routing is the application of improved bio-inspired algorithms in deriving the path towards the destination. Current routing trends determine the optimal path based on throughput, delay or combination of these two metric. Here in this paper, we compare the performance of the two algorithms, AntHocNet (AHN) and Dynamic Source Routing (DSR) for determining the routing path. We introduce the goodput factor as one of the metric for determining the path. Goodput is the number of useful information bits, delivered by the network to a certain destination per unit time .The results show that AHN outperformed DSR in terms of end-to-end delay and overhead. Also the simulation results showed that overhead of AHN was 20.67% lesser than DSR.

*Keywords*: MANET, Ad-Hoc network, AntHocNet, Dynamic Source Routing, Goodput

## 1. Introduction

Computer networks in general have been performing the tasks that were defined by user requirements. With many specialized tasks being introduced, the network has developed a best effort approach in performance by providing QoS. This model has served well till date, but as the end-user applications augments their requirements and a set of similar scenarios repeating in the network, the network tends to increase its complexity and complicate itself in order to deliver the quality in service to the end user. This aspect has generated lot of research interest in active networks. The main objective of this new research field has been to enable the network to evolve to be intelligent. Routing is a process of forwarding the data from a known source to a desired destination. In this process, the data may travel through several intermediate paths, where there is a need to select the best possible optimal nodes to forward the data. This selection of nodes is done to achieve a high performance of the network. There are many existing work done in this area of route discovery which are discussed in the literature by Gelenbe [1], Viswanatha [2] and Nair [3]. The existing algorithms have found the optimal paths considering either one or two QoS parameters or hop counts or cost as the deciding factor for route selection. The objective of this paper is to find an efficient solution for end-to-end delivery which involves performance analysis of the two algorithms and emphasize on bio-inspired algorithm. However, we handle several layers of routers to prove AntHocNet based selection of channels which can be used in route determination. In the recent past, genetic algorithm has found its application in route selection algorithms as discussed in [2]. The fact that AntHocNet learns with the system and applies it to the environment makes it better than the existing algorithms. Thus AHN has been used for path selection in the ever changing network scenario. This will lead to better network performance as compared to the existing predetermined routing.

In today's network routing the challenges arise from the dynamic and unplanned nature of ad hoc network as it is inherent unreliability of wireless communication. A primary challenge of MANET is the design of effective routing algorithm that can adapt its behavior to frequent and rapid changes in the ad hoc network. The main purpose of this work is to evaluate and compare two routing protocols, namely, AntHocNet (AHN) and Dynamic Source Routing (DSR) for wireless ad-hoc networks. The aim of this paper is to show the use of such biologically inspired agents like ant to effectively route the packets in Internet.

The paper is organized as follows: Section 2 addresses the work carried out in this field. Section 3 gives the background work on AHN and DSR. The simulation results are shown in Section 4, and the conclusions and future works are dealt with in section 5.

## 2. Related Work

Routing algorithms play an important role in the path determination process. A good routing algorithm should be able to find an optimal path and it must be simple, stable, converge rapidly and must remain flexible. There exists a lot of routing algorithm which have been developed for specific kind of network as well as for general routing purpose. The existing algorithms are either, table driven or demand-driven protocols. Few of the aspects of routing algorithm are discussed in this section.

### 2.1 Mobile AdHoc Networking

In mobile ad-hoc networks where there is no infrastructure support and nodes being out of range of a source node transmitting packets; a routing procedure is always needed to find a path so as to forward the packets appropriately between the source and the destination. Within a cell, a base station can reach all mobile nodes without routing via broadcast in common wireless networks.

MANET is a collection of nodes, which have the possibility to connect on a wireless medium and form an arbitrary and dynamic network with wireless links. This means that links between the nodes can change with time, new nodes can join the network, and other nodes can leave it. MANET has no permanent infrastructure as it is highly dynamic. Here all mobile nodes act as mobile routers. In literature there exists a large family of ad hoc routing protocols [4]. But there is a need for new routing protocols for specific mobile Ad hoc networks.

### 2.2 Ant Colony Optimization Techniques

Ant Colony Optimization (ACO) is a population-based, general search technique for the solution of difficult combinatorial problems which is inspired by the pheromone trail laying behavior of real ant colonies. ACO is a class of algorithms, whose first member, called Ant System, was initially proposed by Dorigo [8]. The main underlying idea, inspired by the behavior of real ants, is that of a parallel search over several constructive computational threads based on local problem data and on a dynamic memory structure containing information on the quality of previously obtained result.

The main job of ants is to find the shortest path between their nest and a food source. For this reason they use what is known as pheromone. Pheromone is a volatile chemical substance which is secreted by ants. The use of pheromone is an example of indirect communication. The whole process of finding the shortest path is known as Stigmergy, i.e., communicating through the local adaptations of the environment.

Initially one ant goes in search of food from the nest to the food source. On its way it deposits pheromone content. As a result of this a path from nest to the food source is created which resembles a network path in real world. The ants act as mobile agents in the network path. During the next cycle all ants take the path of the pheromone which is deposited by the first ant. On their way these ants also deposit their pheromone. As a result multipath to the destination is created with pheromone deposited. The goal is to determine the optimal path which is dependent on the pheromone content. The path with the highest pheromone content is chosen as the optimal path. Higher pheromone content accounts for higher bandwidth in that path. An optimal path is obtained based on the bandwidth parameter. The selected path will be followed by all ants. This leads to the derivation of the optimal path.

Ant Colony Based Routing Algorithm (ARA) was proposed to reduce overhead, mainly because routing tables are not interchanged among nodes [5]. It consist of three phases namely
- Route Discovery phase
- Route Maintenance
- Route Failure Handling.

The Route Discovery phase has two mobile agents i.e. Forward Ant (FANT) for route request and Backward Ant (BANT) for route reply to create new routes. The FANT packets have unique sequence number and source address is broadcasted by the sender and will be passing on by the neighbours of the sender. Node receiving the FANT for the first time generates a record with entries of destination address (Source address of FANT), next hop (address of previous node), and pheromone value (number of hops the FANT needed to reach this node). The destination node extracts information of FANT, destroys it and creates BANT which establish pheromone track to destination node.

In Route Maintenance phase, DUPLICATE ERROR flag is set for duplicate packets to prevent from looping problems. ARA also allows for the evaporation of pheromone by decrementing pheromone factor in route table.

In Route Failure Handling phase, node deactivates the path by reducing pheromone value to 0 in corresponding route table entry and go to the Route Discovery phase for selecting path and sending packets to the destination over that path [6-7].

### 3.   AntHocNet and DSR

AntHocNet is the hybrid algorithm which includes the concept of Ad Hoc networks and AntNet [8-9]. During reactive path setup, multiple routes are set on demand by broadcasting reactive FANT and gather information about quality of path. In case of broadcasting the nodes receives number of ants. Here the node compares the path travelled by each new ant to that of former received ants of this generation. It rebroadcasts only if its number of hops and travel time are both within an acceptance factor of best forward ant. Once paths are setup, source starts sending proactive FANT to destination on the basis of pheromone values combined with small probability value at each node which has been broadcasted. Fig. 1 depicts the actual follow of the working principle of AHN.

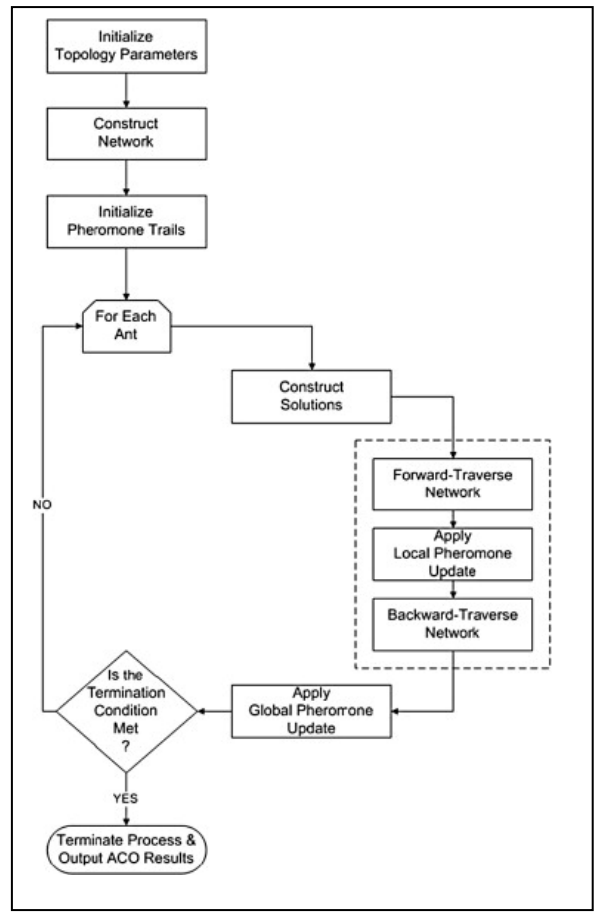

Fig. 1 AntHocNet

Dynamic Source Routing (DSR) is a routing protocol for wireless mesh networks. This protocol is truly based on source routing whereby all the routing information is maintained (continually updated) at mobile nodes. It has only two major phases, which are Route Discovery and Route Maintenance. Route Reply would only be generated if the message has reached the intended destination node (route record which is initially contained in Route Request would be inserted into the Route Reply).

The DSR protocol is composed of two mechanisms that work together to allow the discovery and maintenance of source routes in the ad hoc network:

i)   Route Discovery is the mechanism by which a node S wishing to send a packet to a destination node D obtains a source route to D. Route Discovery is used only when S attempts to send a packet to D and S has no prior knowledge about the route to D.

ii)  Route Maintenance is the mechanism by which node S is able to detect, while using a source route to D. If the route to D is no longer used it indicates that the network topology has changed or the link along the route no longer exists. When Route Maintenance indicates a source route is broken, S can attempt to use any other route it happens to know to D, or can invoke Route Discovery to find again a new route. Route Maintenance is used only when S is actually sending packets to D [10].

### 4. Simulation Results

The simulation has been carried out on NS2 [11]. The following table show the initial setup assumption made for the simulation run. The result analysis is tested on end-to-end delay metric, throughput, goodput and packet delivery ratio. Table-I depicts the details of the configuration setup parameters.

Table-I. Parameters setup for node configuration

| Parameters | Settings |
|---|---|
| Simulation duration | 180 seconds |
| Moving range | 2500*1500 meters |
| Business type | FTP |
| Communication radium | 250 meters |
| Network bandwidth | 2 Mbps |

The Fig. 2, depicts one of the simulation run's output with 32 nodes. The simulation is carried out for 180 seconds and the results are derived from the trace file.

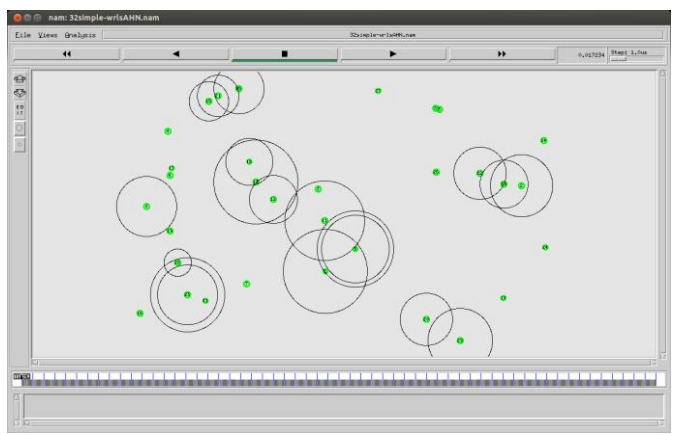

Fig. 2  Simulation setup for AntHocNet

Table-II indicates the mobile node configuration setting which has been used for the AntHocNet routing protocol. The initial energy chosen is 100joules. A same setting has been used by the DSR algorithm except that routing protocol needs to be changed to DSR.

Table-II. Parameters setup for mobile node configuration

| Parameters | Settings |
|---|---|
| val(channel) | Channel/WirelessChannel |
| val(prop) | Propagation/TwoRayGround |
| val(mac) | Mac/802_11 |
| val(nn) | 16 |
| val(ifq) | CMUPriQueue |
| val(rp) | AntHocNet |
| val(netif) | Phy/WirelessPhy |
| val(ifqlen) | 50 |
| set opt(energymodel) | EnergyModel |

Trace files are post-processed to calculate the delay of each transmitted packet during the simulation. Departure time, packet ID, delay and number of hops are written in a so-called plot file. Average delay is calculated dividing the total delay by the number of packets arrived at destination.

Energy utilisation is defined as the amount of energy required to deliver the data packets from source to destination. The results were got based on the energy required for a set of nodes. We observe that initially when the number of nodes is less the energy utilised will be almost same for both AntHocNet and DSR protocols. But as the number of nodes increases or to be more specific goes above 64, the energy required for packet delivery is more in DSR compared to AntHocNet.

Table-III. End-to-end delay analysis

| Number of Nodes | AntHocNet | Dynamic Source Routing |
|:---:|:---:|:---:|
| 16 | 136.836 | 134.962 |
| 32 | 147.257 | 259.435 |
| 50 | 204.443 | 313.918 |
| 64 | 287.857 | 335.351 |
| 80 | 220.758 | 398.248 |
| 128 | 730.089 | 886.682 |

Based on the above results, AntHocNet is found to have lesser end-to-end delay than DSR for the same topology.

Table-IV Throughput and goodput of AHN and DSR

| No. of Nodes | Throughput (kbps) | | Goodput (kbps) | |
|:---:|:---:|:---:|:---:|:---:|
| | AntHocNet | DSR | AntHocNet | DSR |
| 16 | 710.24 | 523.24 | 658.78 | 485.33 |
| 32 | 737.72 | 767.26 | 684.26 | 711.66 |
| 50 | 452.70 | 524.75 | 419.89 | 486.67 |
| 64 | 234.80 | 424.27 | 217.79 | 393.50 |
| 80 | 813.58 | 1123.76 | 754.63 | 1042.12 |
| 100 | 682.77 | 839.29 | 633.29 | 778.33 |
| 128 | 127.77 | 524.87 | 118.52 | 486.45 |

Throughput is the average rate of successful message delivery over a communication channel. This data may be delivered over a physical or logical link, or pass through a certain network node. The throughput is usually measured in bits per second (bit/s or bps), and sometimes in data packets per second. It is essentially synonymous to digital bandwidth consumption. Table-IV describes the throughput and goodput values obtained for different simulation run.

Goodput is the application level throughput, i.e. the number of useful information bits, delivered by the network to a certain destination, per unit of time. The amount of data considered excludes protocol overhead bits as well as retransmitted data packets. This is related to the amount of time from the first bit of the first packet is sent (or delivered) until the last bit of the last packet is delivered. The goodput factor derived for both the algorithm is as shown in Fig. 3.

Overhead is the difference between the throughput and goodput. Average overhead of AHN and DSR is calculated across the simulation run with different node numbers and it is seen that AHN's overhead is 20.67% lesser than DSR.

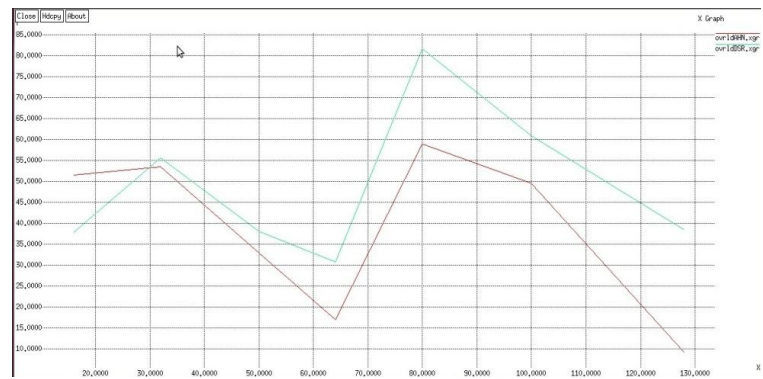

Fig. 3 comparison of goodput of AHN and DSR

Though  AntHocNet utilizes more bandwidth than DSR as the number of nodes in the simulation is increased,  AntHocNet has lesser overhead than DSR.

The data packet delivery ratio is defined as the number of successfully delivered data packets to the number of data packets generated by the source. Packet Delivery Ratio trace files are post-processed to calculate the delivery ratio of data packets. That is, the relation between sent packets and received packets. As usual, the performance of AntHocNet and DSR were considered. It was observed that AntNet is relatively consistent and stable as compared to DSR. This result is observed when traffic in the network does not exist. It is observed that both the protocols, the Packet Delivery Ratio decreases with the increase in error rate, or correspondingly the packet loss. Furthermore, it was observed that when the packet loss is less, DSR gives the best Packet Delivery Ratio.

Energy of each node is set to 100 joules. As simulation progresses, nodes lose energy more frequently in AHN than in DSR. This happens because in AHN, there are extra packets such as ant packets (proactive forward and backward ants and reactive forward and backward ants) that are used for route discovery and maintenance. Bandwidth and other network parameters such as energy, and bandwidth will be utilized for the purpose of these ants. This reduces the resources for normal data packets and leads to lesser PDF in AHN than in DSR as shown in the Fig. 4.

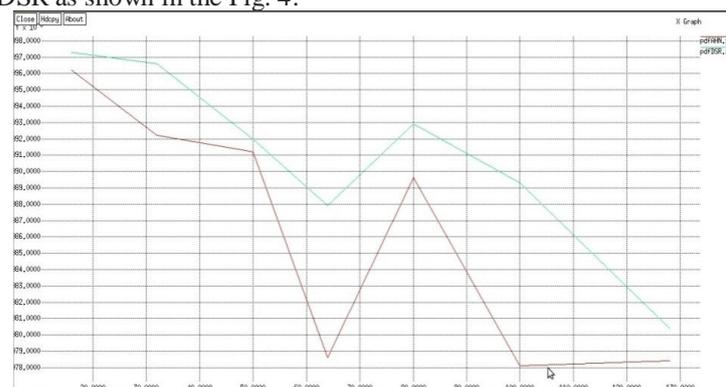

Fig. 4 comparison of PDF of AHN and DSR

The above results show that AntHocNet has lesser packet delivery ratio than DSR which means fewer numbers of packets get delivered to the destination nodes than in DSR. This is due to the fact that the simulation was carried out for only 180 second. The packets will definitely reach the destination if the simulation time is increased. It was observed that the data packet delivery in AHN takes some time as ant packets are large in size and consumes bandwidth.

## 5.  Conclusion

This paper simulates and analyzes different topological node structures on NS2 environment, comparing the performance of AntHocNet and DSR. It was observed that the metric like end-to-end delay, throughput and goodput showed better performance in AHN than in DSR as tabulated in the result. While the packet delivery ratio was less in AHN than in DSR as ant agents occupy certain bandwidth and simulation setup time considered was only 180 seconds.

The paper can be extended further by comparing the other bio inspired algorithm where learning approaches have been performed like ants. A comparative analysis of learning overhead in order to improve the optimal path determination, can be extended as the future work.

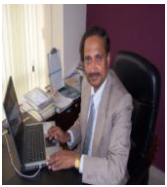

**T.R.Gopalakrishnan Nair** holds M.Tech. (IISc, Bangalore) and Ph.D. degree in Computer Science. He has 3 decades experience in Computer Science and Engineering through research, industry and education. He has published several papers and holds patents in multi domains. He is winner of PARAM Award for technology innovation. Currently he is the Saudi Aramco Endowed Chair, Technology and Information Management, PMU, KSA and VP of Research and Industry in Dayananda Sagar Institutions, Bangalore, (Sabbatical), India.

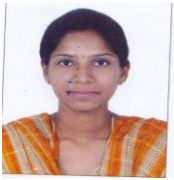

**Kavitha Sooda** holds M.Tech in Computer Scinece and Engineering. She has eleven years of teaching experience and pursuing her Ph. D from JNTU, Hyderbad. Her interest includes Routing techniques, QoS application, Cognitive networks and Evolutionary algorithms. She has twelve papers based on her research area. Currently she works as Assistant Professor at Nitte Meenakshi Institute of Technology, Bangalore, India.

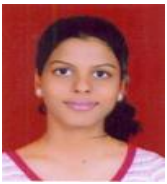

**Kavya B S** did her M.Tech., in Network and Internet Engineering(NIE) in the year 2010. Currently working as lecturer in the department of CSE at Nitte Meenakshi Institute of Technology, Bangalore. Her area of interest is Networking

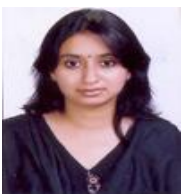

**Sushma M** did her M.Tech., in Network and Internet Engineering(NIE) in the year 2010. Currently working as lecturer in the department of CSE at Nitte Meenakshi Institute of Technology, Bangalore. Her area of interest includes Networking and cryptography.